\documentclass[aps,prb,twocolumn,groupedaddress,showpacs]{revtex4}

\usepackage{graphicx}

\begin{document}

\title{The structure of binary Lennard-Jones clusters: The effects of 
atomic size ratio}

\author{Jonathan P.~K.~Doye}
\affiliation{Department of Chemistry, University of Cambridge, Lensfield Road, Cambridge CB2 1EW, United Kingdom} 
\author{Lars Meyer}
\affiliation{Department of Chemistry, University of Cambridge, Lensfield Road, Cambridge CB2 1EW, United Kingdom} 

\date{\today}

\begin{abstract}
We introduce a global optimization approach for binary clusters that for a 
given cluster size is able to directly search for the structure and 
composition that has the greatest stability. We apply this approach to 
binary Lennard-Jones clusters, where the strength of the 
interactions between the two atom types is the same, but where the atoms
have different sizes. We map out how the most stable structure depends
on the cluster size and the atomic size ratio for clusters with up 
to 100 atoms and up to 30\% difference in atom size. A substantial
portion of this parameter space is occupied by structures that are 
polytetrahedral, both those that are polyicosahedral and those that  
involve disclination lines. 
Such structures involve substantial strains for one-component 
Lennard-Jones clusters, but can be stabilized by the different-sized atoms 
in the binary clusters. These structures often have a `core-shell' geometry,
where the larger atoms are on the surface, and the smaller atoms are in 
the core.
\end{abstract}


\maketitle

\section{\label{sect:intro}Introduction}

There has been much recent interest in binary clusters, both from a
fundamental and a technological perspective. For example, alloy clusters
are of particular importance, because of their potential catalytic 
properties.\cite{Lee95,Molenbroek98} 
Furthermore, binary clusters offer the opportunity to 
tailor their properties through the choice of atom types and composition, 
potentially leading to new behaviour that is not possible for
single component clusters. 
Here our focus is on the structure of binary clusters, which is one of 
their most important properties, and is a prerequisite for understanding 
many of the other properties of such clusters.

For monoatomic clusters, Lennard-Jones (LJ) clusters provide a 
well-characterized model system, for which the effects of a cluster's
finite size on the structure, thermodynamic and dynamic properties are
well understood. For example, putative global minima 
are now available for all LJ clusters with up to 1600 atoms,
\cite{Northby87,WalesD97,Romero99,Xiang04,Xiang04b,Shao05} 
and the size evolution of the structure of larger clusters is well 
understood.\cite{Raoult89a,Doye01b,Doye02b}

For binary clusters, there is a similar need for an archetypal system 
to understand the effects that can control the structure of a binary
cluster. We propose that binary Lennard-Jones (BLJ) clusters could
provide just such a model system, and this paper aims to 
start the systematic exploration of this model.
One advantage of using BLJ clusters as a model system is the relative 
simplicity of the potential. There are only four effective
parameters that characterize the interactions between the 
different combinations of atoms, making it possible to 
systematically study the structure of BLJ clusters
as a function of these parameters. By contrast, even the
simplest many-body metal potential would have a considerably 
greater number of parameters, and so it only becomes feasible
to study a series of example binary metal systems,
which just represent a set of points in this state space of potentials.

Binary clusters also offer considerable additional challenges to the 
theoretician, compared to the one-component case. 
Firstly, for a given cluster, there
are many more minima on the potential energy surface, 
because of the presence of ``homotops'',\cite{Jellinek96} isomers with
the same geometric structure, but which differ in the labelling of the atoms.
Secondly, the composition provides an 
additional variable to consider.
For example, 
the task of obtaining 
the lowest-energy structures
for all compositions and all sizes up to 100 atoms would 
require 5050 different global minima to be found.
Instead, most studies have either just considered a few sizes and
explored how the structure depends on composition,
\cite{Rey96b,Calvo04,Rossi04,Rapallo05,Rossi05}
or kept the composition fixed and studied the size 
dependence.\cite{Massen02,Bailey03}
By doing such a selective survey, there is the possibility
that the most interesting and stable structures for the systems are missed.

Here, we take a different approach, neither trying to find the
global minima for every size and composition, nor taking an arbitrary cut
through this space. 
Since one is usually just interested in particularly stable structures, 
finding all the global minima is unnecessary, 
and so we instead directly search for these particularly stable structures. 
In particular, we use the composition as a variable in our
global optimization and for a given size we attempt to find the 
most stable composition.

\begin{figure}
\includegraphics[width=6.4cm]{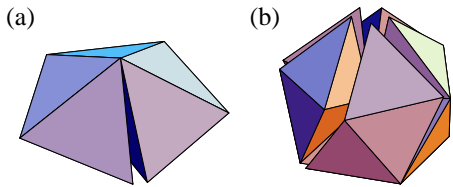}
\caption{\label{fig:gaps} (Colour online)
Examples of the strain involved in packing tetrahedra. 
(a) Five regular tetrahedra around a common edge produce a gap of $7.36^\circ$.
(b) Twenty regular tetrahedra about a common vertex produce gaps equivalent 
to a solid angle of 1.54 steradians.}
\end{figure}

Our focus in this paper is to use BLJ clusters to understand how
different types of cluster structure can be stabilized purely through the
two atom types being of different size. In particular, we are interested 
in polytetrahedral structures,\cite{NelsonS} for which all the occupied space 
can be divided up into tetrahedra with atoms at their corners. 
For a one-component system such packings are said to be `frustrated',
because regular tetrahedra do not tesselate, 
as illustrated in Fig.\ \ref{fig:gaps}. The best local packing of 
tetrahedra involves packing five tetrahedra around a common edge, 
but with regular tetrahedra there is a small angular deficit. 
As larger structures are made up of regular tetrahedra, these gaps grow 
rapidly in size.
So for a 13-atom icosahedron, which can be thought of as a packing of 20 
slightly irregular tetrahedra around a common vertex, the distance between
adjacent atoms on the surface is 5.15\% longer than that between the 
central atom and a surface atom. For monoatomic systems, there
will be an energetic penalty associated with this strain. 
However, the associated strain can be completely removed in a binary system
simply by choosing the central atom to be 9.79\% smaller. 
Similarly, Frank-Kasper phases,
bulk polytetrahedral crystals, are only found for 
alloys.\cite{FrankK58,FrankK59,Shoemaker}
Furthermore, stabilization of polytetrahedral structures in 
binary metal clusters in which there is a size mismatch between the two 
atom types has previously been seen.\cite{Rossi04,Rapallo05,Rossi05}

In section \ref{sect:methods} we describe the choice of potential parameters,
how we analyse the energetics, and our global optimization approach.
In section \ref{sect:fixed_gmin}, for three cluster sizes we present case 
studies of how the lowest-energy structures depend on both the composition and 
the atomic size ratio, and in section \ref{sect:vary_gmin} we present our
systematic survey of the optimal structures and compositions for all clusters
with up to 100 atoms and up to 30\% difference in atom size. 
A brief report of some of the work presented here has appeared previously.\cite{Doye05d}

\section{\label{sect:methods}Methods}

\subsection{\label{subsect:potential}Potential}

Here, we use a binary Lennard-Jones (BLJ) potential:
\begin{equation}
E=4\sum_{i<j} \epsilon_{\alpha\beta} \left[ 
              \left({\sigma_{\alpha\beta}\over r_{ij}}\right)^{12}-
              \left({\sigma_{\alpha\beta}\over r_{ij}}\right)^6
                                    \right], 
\end{equation}
where $\alpha$ and $\beta$ are the atom types of atoms $i$ and $j$,
and $\epsilon_{\alpha\beta}$ and $2^{1/6}\sigma_{\alpha\beta}$
are the pair well depth and equilibrium pair separation, respectively, 
for the interaction between atoms $i$ and $j$. In its most general form
the BLJ potential has four effective parameters, namely 
$\epsilon_{AB}$, $\epsilon_{BB}$, $\sigma_{AB}$ and $\sigma_{BB}$, if 
$\epsilon_{AA}$ and $\sigma_{AA}$ are used as the units of energy and length,
respectively. 
Here, as we wish to consider the effects
of purely the size ratio on the most stable structures, we choose 
$\epsilon_{AA}=\epsilon_{AB}=\epsilon_{BB}=\epsilon$.
Furthermore, as we define $\sigma_{AB}$ using the Lorentz rule,
$\sigma_{AB}=(\sigma_{AA}+\sigma_{BB})/2$, 
in this case there is effectively just one parameter in the potential, 
namely the size ratio of the two atoms, $\sigma_{BB}/\sigma_{AA}$.

Initially, we thought of directly searching for 
the most stable composition and size ratio for a cluster of a particularly size
in our global optimization runs.
However, we quickly found that increasing the size disparity between the 
two types of particles leads to a virtual monotonic decrease in the energy. 
Consequently, our optimization runs led to structures with huge 
differences in the sizes of the atoms. 
These structures consist of a core
of tiny atoms surrounded by a shell of large atoms, where the large atoms are 
able to interact strongly with all the atoms in core. They are clearly 
unphysical and so this approach was abandoned in favour of using only the
composition as a variable during the global optimization, and considering 
different size ratios independently. 

We chose to look at the structures for size
ratios in the range $1<\sigma_{BB}/\sigma_{AA}<1.3$, as we wished to see
if polytetrahedral structures would be stabilized as we move away from the
one-component Lennard-Jones reference system (i.e.\ 
$\sigma_{BB}/\sigma_{AA}=1$).   
Due to the symmetry of the energetic interactions, 
exactly the same structural behaviour will be seen in
the parameter range $0.769<\sigma_{BB}/\sigma_{AA}<1$, 
except that the role of the A and B atoms will be reversed. 
Six values were considered; namely, $\sigma_{BB}/\sigma_{AA}=1.05$, 
1.1, 1.15, 1.2, 1.25 and 1.3.

There has been a certain amount of previous work on binary LJ clusters,
as a general model 
for binary systems,\cite{Garzon89,Clarke93,Clarke94,Rey95,Cozzini96}
as a model of binary rare-gas 
clusters\cite{Robertson88,Frantz97,Munro02,Calvo04,Cleary06}
and as a playground for testing how the potential parameters could
be used to tailor the cluster's structural 
properties.\cite{Sabo03,Sabo04a,Sabo04b,Cleary06}
Of these studies, a few have considered the same set of 
parameters as we have here,\cite{Garzon89,Clarke94,Cozzini96} with some 
tendencies to form core-shell clusters noted.\cite{Garzon89,Clarke94} 
However, nothing like the systematic survey presented here has
been tried previously.

To apply the current approach, one needs to be able to compare
the relative stabilities of clusters with the same number of atoms 
but different compositions. For the current system, as all the $\epsilon_{ij}$
are the same, the most natural way is to compare their absolute energies
directly. 

For systems, where the energetic interactions are more varied, 
this issue becomes more subtle. 
There are a number of approaches one might take depending on 
to what the `stability' is being measured with respect.
For example, a simple approach would be to compare the energies to a linear
interpolation between the energies of the pure systems. In bulk, 
this would amount to comparing the stability with respect to separation
into the pure phases. For clusters, it would be sensible to somewhat amend
this approach, as it would automatically put the pure clusters on the same 
energetic footing, irrespective of whether they represented 
particularly stable sizes (or not) for the pure systems. For instance,
one could instead measure the energies with respect to a linear interpolation 
between smoothly varying functions, call them $E_{\rm ave}^\alpha(N)$, 
that captured the general size dependence of the energies of the pure clusters.

Alternatively, one might be interested in maximizing the 
stability of the cluster with respect to the liquid state, i.e.\ finding
the composition with the maximum melting point.
For example, for systems where the cross-terms favour mixing, 
there will be a stabilization of intermediate compositions for 
both the solid and the liquid, and so a non-linear interpolation 
would be required to detect enhanced thermostability from the
background mixing effects.

In order to understand better how particular structures are stabilized,
it is useful to decompose the energy into different terms. 
Firstly, we divide the energies into contributions from those pairs
of atoms that are nearest neighbours from those that are not, i.e.
\begin{equation}
E=E_{\rm nn}+ E_{\rm nnn}
\end{equation}
where $E_{\rm nn}$ and $E_{\rm nnn}$ are the energetic contributions
from nearest neighbours and non-nearest neighbours. Nearest neighbours are
simply defined using a distance criterion. In practice we used 1.3 times
the relevant equilibrium pair separation, 
but as there is usually a clear separation between
nearest-neighbour and next-neighbour shells for the ordered structures that we 
find to be the global minima, the precise value is not critical.

Secondly, we divide up $E_{\rm nn}$ by defining 
a strain energy in terms of the difference in
energy between the nearest-neighbour energy, and the energy this term 
would have if all the nearest neighbours had the same separation (measured
with respect to the relevant equilibrium pair separation): i.e.\
\begin{equation}
E_{\rm nn}=n_{\rm nn} V_{LJ}(r_{\rm nn}' \sigma)+E_{\rm strain},
\end{equation}
where $n_{\rm nn}$ is the number of nearest neighbours, and
$r_{\rm nn}'$,
the reduced average separation of these nearest neighbours is defined as
$\langle r_{ij}/\sigma_{\alpha\beta}\rangle_{\rm nn}$, where
the average is taken over only those pairs of atoms that 
are nearest neighbours.\cite{prevdef}
Hence, $E_{\rm strain}$ measures the energetic penalty
arising because a structure has a distribution of nearest-neighbour 
distances. 

As $E_{\rm nnn}$ is relatively insensitive to structure, the 
best structure is determined mainly by the balance between maximizing 
$n_{\rm nn}$ and minimizing $E_{\rm strain}$. For one-component
polytetrahedral clusters, both $n_{\rm nn}$ and $E_{\rm strain}$ are 
usually large, 
and by following $E_{\rm strain}$ as a function of $\sigma_{BB}/\sigma_{AA}$ 
will allow us to see how this strain can be relieved by having different
sized atoms.

\subsection{\label{subsect:GO}Global optimization}

The basin-hopping global optimization algorithm\cite{WalesD97,Li87a}
has proved to be very successful for the optimization of a wide range
of clusters. The method simply involves doing a constant 
temperature Metropolis Monte Carlo 
simulation, where after each step a local minimization is performed on
the resulting configuration, and the acceptance criterion is based upon
the energies of these minimized configurations. The success of the method
is because it effectively searches a transformed potential energy surface,
for which the dynamics and thermodynamics is more helpful for 
optimization.\cite{Doye98a,Doye98e}

In addition to the standard basin-hopping moves (random displacements
of all atoms and rotations of low-energy atoms around the centre of mass)
used for one-component systems, for binary clusters
it is important to incorporate moves into the algorithm that allow 
efficient exploration of the space of homotops and composition. 
The two moves that we used involved swapping the identities of an A 
and a B atom,\cite{Calvo04} and changing the identity of a single atom.
In optimizations runs where the composition was held fixed, only the former
composition-preserving moves were used. Typically, these types of moves
represented 50\% or more of the total moves.

The current optimization task of finding the structure and composition 
of a binary cluster that has the lowest energy 
(we shall call this the `compositional global minimum') is very challenging, 
because the number of possible structures increases very rapidly with size.
Firstly, the number of geometric isomers scale as 
$\exp(\alpha N)$.\cite{Tsai93a,Still99}
Secondly, the number of possible homotops for a given geometric structure
is proportional to 
${N \choose N_A}$
(assuming that all the homotops are locally stable).
Thirdly, the number of possible compositions increases linearly with $N$.
For example, for LJ$_{100}$ the estimated number of geometric isomers
is of the order of $10^{39}$. If all these are stable for the binary clusters
and all homotops are possible, this then gives an estimate
of $10^{69}$ energetically different minima in the space that we search 
to find the compositional global minimum of BLJ$_{100}$. 
It is, therefore, important that we make as much use of the structural
information gleaned from unbiased runs to aid the optimization task.

Our initial strategy was to perform an initial series of runs for each 
cluster size from random starting points at a moderate temperature, similar
to that typically used for one-component LJ clusters. 
Subsequently, we performed low temperature runs from the best structures
found from the above runs. 
The rationale is that for $\sigma_{BB}/\sigma_{AA}$ values sufficiently 
close to one, the energy differences between homotops can be significantly
smaller than that between geometric structures. 
For example, this feature manifests 
itself in the thermodynamics of such binary clusters as low-temperature 
heat capacity peaks associated with permutational 
disordering.\cite{Lopez93,Frantz97}
The hope is then that the high temperature runs would find the best 
geometric structures, and the low temperature run the best homotop 
consistent with this geometric structure.
At larger $\sigma_{BB}/\sigma_{AA}$, this separation of energy scales 
breaks down, and so the low-temperature runs are less important.
We also performed series of runs, where the best structure for a particular
size and size ratio were used as starting points for basin-hopping runs
at different size ratios and sizes (with the requisite number of atoms
added or removed). These different types of runs were applied iteratively
until convergence seemed to be reached.

\section{\label{sect:results}Results}

\subsection{\label{sect:fixed_gmin}Case studies}
Before we present results whilst directly searching for the global minima 
in both composition and configuration space, it is useful to get a feel
for the energetics as a function of composition.
To do this, we map out the landscape associated with the energy of the
global minima as a function of $n_A$ and $\sigma_{BB}/\sigma_{AA}$ for a
number of examples, namely
BLJ$_{13}$, BLJ$_{45}$ and BLJ$_{55}$.
For the latter two, this represents a significant effort, since we have 
found the global minima at 1320 and 1620 different points in this space,
respectively.

\begin{figure}
\includegraphics[width=8.4cm]{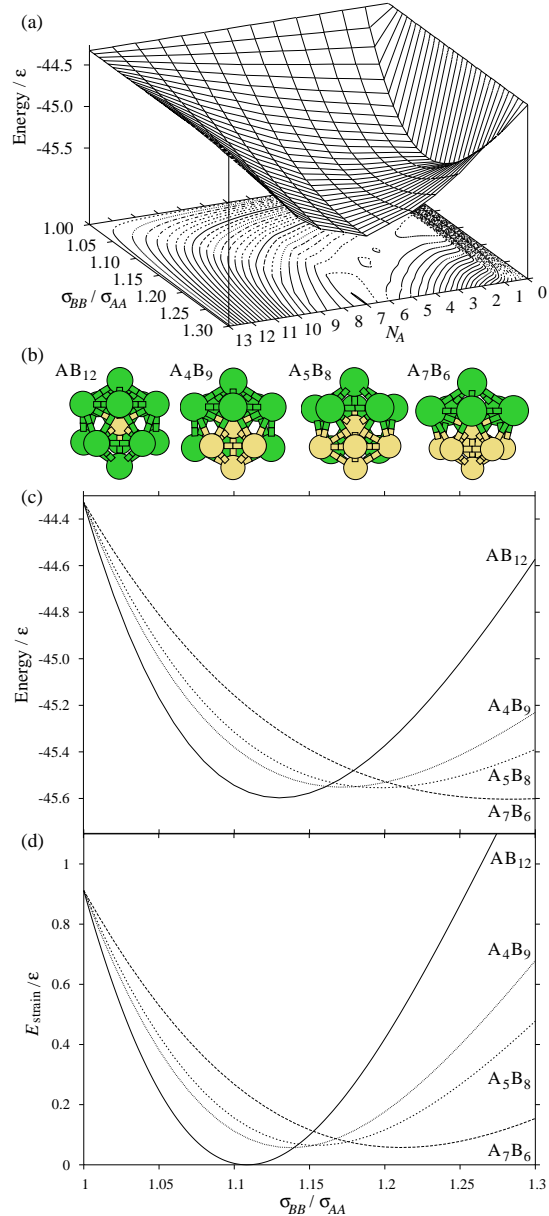}
\caption{\label{fig:BLJ13} (Colour online) BLJ$_{13}$.
(a) Landscape showing the dependence of the energy of the global minimum
on $\sigma_{BB}/\sigma_{AA}$ and $N_A$.
(b) The four compositional global minima and the dependence of 
(c) their total energies and (c) strain energies 
on $\sigma_{BB}/\sigma_{AA}$.}
\end{figure}

For BLJ$_{13}$, the structure of the global minima is always the centred 
icosahedron, so Fig.\ \ref{fig:BLJ13}(a) simply represents the variation
of the energy of this one structure.
Close to $\sigma_{BB}/\sigma_{AA}=1$, as expected, the structure with 
one small atom at the centre of the cluster is the 
most stable composition. The distance between adjacent vertices 
of the perfect icosahedra is 5.15\% longer than that between a vertex
and the centre.
Consistent with this, the strain energy goes to zero when 
$\sigma_{BB}/\sigma_{AB}$ takes this value, i.e.\
$\sigma_{BB}/\sigma_{AA}=1.1085$. The minimum in the total energy 
is displaced to slightly larger $\sigma_{BB}/\sigma_{AA}$, namely 1.1303,
because this leads to a greater contribution to the energy from next neighbours.
Beyond this, the energy of the AB$_{12}$ structure rises, 
and at $\sigma_{BB}/\sigma_{AA}=1.1614$, the A$_4$B$_{9}$ structure 
becomes the most stable composition. With three smaller atoms in the surface, 
this structure is more able to relieve the compressive strain 
that builds up in the AB$_{12}$ structure.
With further increases in $\sigma_{BB}/\sigma_{AA}$ the global minimum changes 
twice more with the number of A atoms increasing in order to achieve
a lower strain energy. These transitions give a flat bottom to the 
landscape in Fig.\ \ref{fig:BLJ13}(a). 
\begin{figure}
\includegraphics[width=6.4cm]{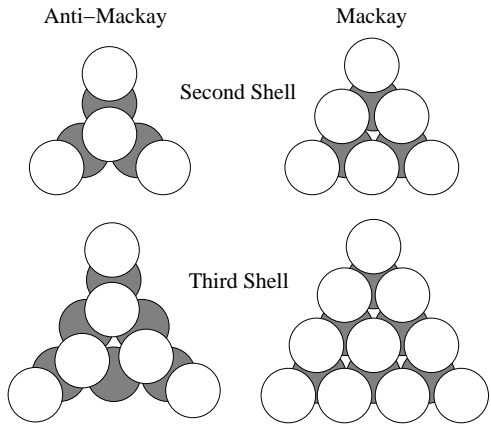}
\caption{\label{fig:layer} The two possible overlayers for growth
around a complete Mackay icosahedron. These are illustrated for 
a single face of the underlying 13-atom (top) and 55-atom (bottom) 
Mackay icosahedron.}
\end{figure}

\begin{figure*}
\includegraphics[width=18cm]{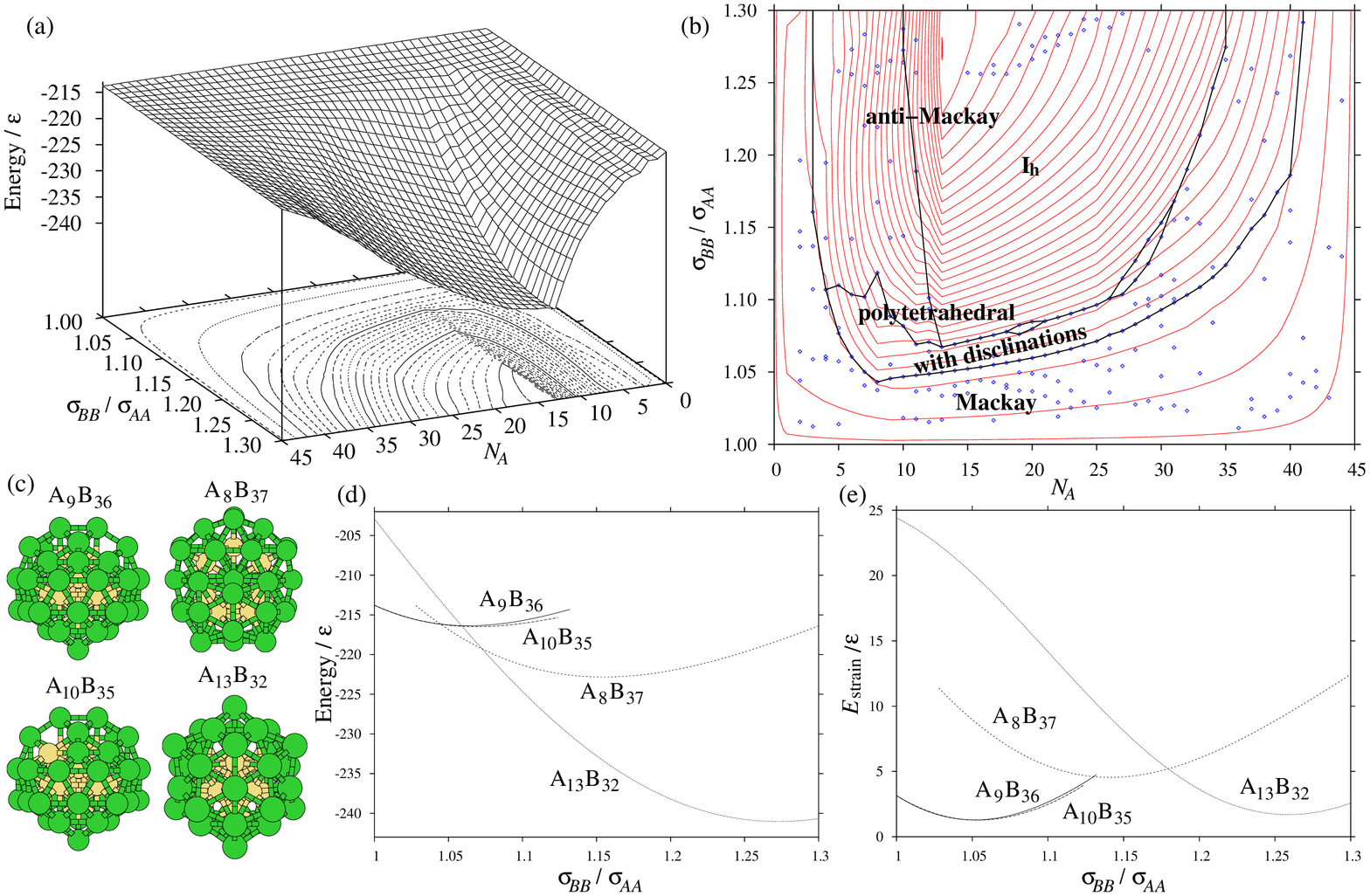}
\caption{\label{fig:BLJ45} (Colour online) BLJ$_{45}$.
(a) Landscape showing the dependence of the energy of the global minimum
on $\sigma_{BB}/\sigma_{AA}$ and $N_A$.
(b) Zero-temperature structural phase diagram showing how the structure of the 
global minimum depends on $\sigma_{BB}/\sigma_{AA}$ and $N_A$, superimposed
on the contour plot of the landscape in (a).
Each data point corresponds to a change in the structure of the global minimum
as a function of $\sigma_{BB}/\sigma_{AA}$. The region where the 
anti-Mackay icosahedron is global minimum is labelled $I_h$.
(c) The four compositional global minima and the dependence of 
(d) their total energies and (e) strain energies 
on $\sigma_{BB}/\sigma_{AA}$.}
\end{figure*}

BLJ$_{45}$ shows a much richer structural behaviour, because
there are a variety of competing geometric structures. 
Firstly, there are two ways that atoms can be added around the
13-atom Mackay icosahedron, as illustrated in Fig.\ \ref{fig:layer}. 
In general, the Mackay overlayer continues
the face-centred-cubic (fcc) packing of the twenty vertex-sharing 
fcc tetrahedra that make up 
a Mackay icosahedron, whereas the anti-Mackay overlayer adds atoms 
in sites that are hcp with respect to the fcc tetrahedra.
The anti-Mackay layer has a lower surface density, and consequently
has a greater number of nearest neighbours, 
but for monoatomic systems also a larger strain energy. Typically,
therefore, growth starts off in the anti-Mackay overlayer, and then switches
to the Mackay overlayer, as the layer grows.

For growth on the 13-atom icosahedron, the anti-Mackay overlayer 
maintains the polytetrahedral character of the structures with each interior
atom having a local icosahedral coordination shell (we call such structures
polyicosahedral), whereas
the Mackay overlayer introduces some octahedral interstices into the 
structure.  For LJ clusters, $N=31$ is the first size at which the global 
minimum adopts the Mackay overlayer.\cite{Northby87}
We chose to examine BLJ$_{45}$ in detail, because at this size the
anti-Mackay overlayer is able to be completed, giving a structure
with point group $I_h$ that we call the anti-Mackay icosahedron.

Secondly, there is the possibility of structures that are polytetrahedral,
but that are not polyicosahedral. Instead, there are interior atoms that
have a coordination number ($Z$) greater than 12. 
For polyicosahedral structures,
there are five tetrahedra around every interior edge.
However, as we noted in the introduction, the corresponding 
packings of regular tetrahedra involve gaps that increase rapidly 
with the size of the packing (Fig.\ \ref{fig:gaps}). 
As a consequence, it is not possible to generate a bulk polyicosahedral 
packing, and for clusters, the surface density decreases
and the average nearest-neighbour distance between the surface atoms increases
as the structures become bigger.

\begin{figure*}
\includegraphics[width=18cm]{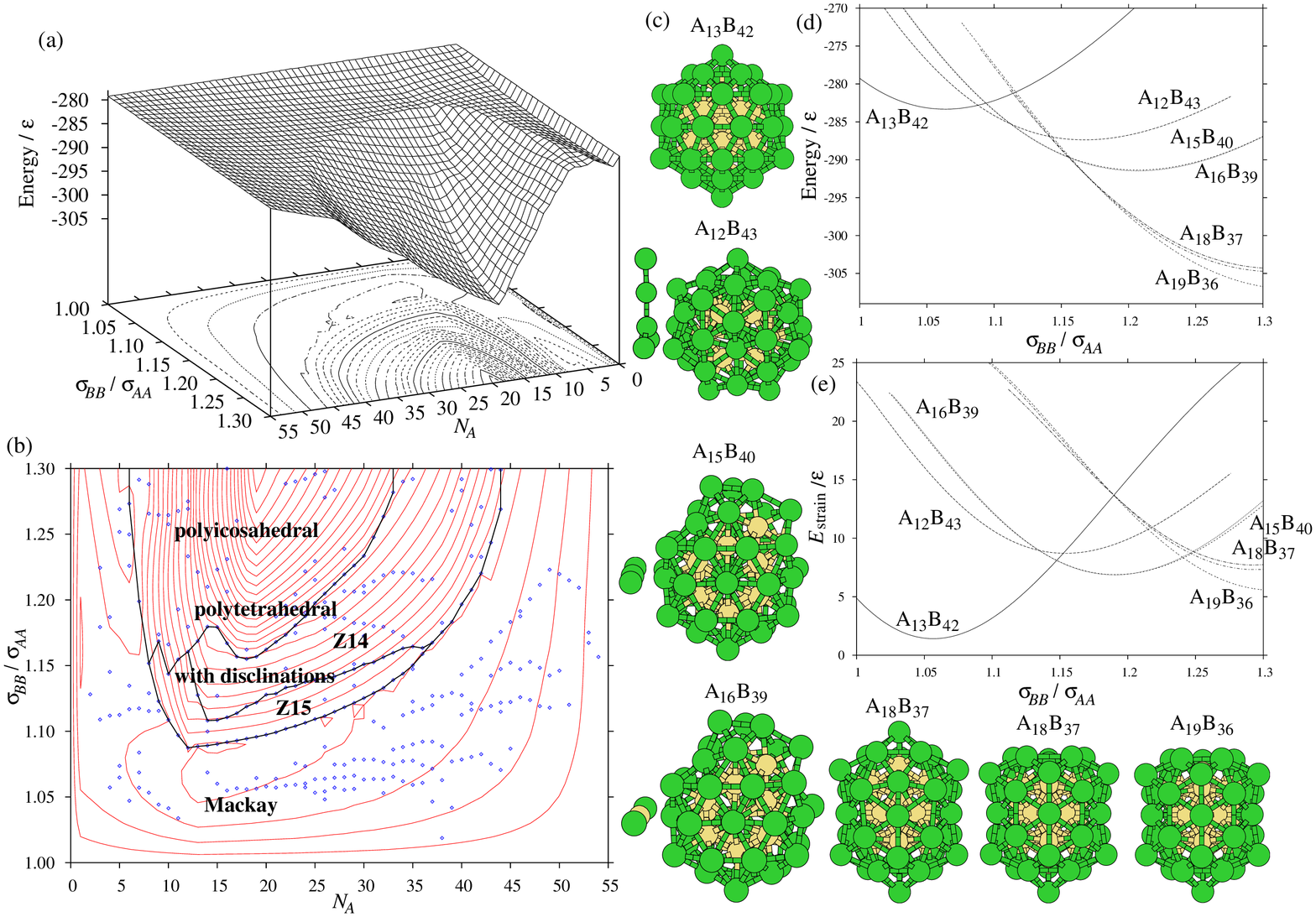}
\caption{\label{fig:BLJ55} (Colour online) BLJ$_{55}$.
(a) Landscape showing the dependence of the energy of the global minimum
on $\sigma_{BB}/\sigma_{AA}$ and $N_A$.
(b) Zero-temperature structural phase diagram showing how the structure of the 
global minimum depends on $\sigma_{BB}/\sigma_{AA}$ and $N_A$, superimposed
on the contour plot of the landscape in (a).
Each data point corresponds to a change in the structure of the global minimum
as a function of $\sigma_{BB}/\sigma_{AA}$. The region where 
polytetrahedral structures involving disclinations are 
the global minima is divided into two subregions, depending 
on whether the highest coordination number for an atom is 14 or 15.
(c) The seven compositional global minima and the dependence of 
(d) their total energies and (e) strain energies 
on $\sigma_{BB}/\sigma_{AA}$.
In (c) for the relevant global minima the disclination networks are depicted
to the left of the structure.}
\end{figure*}

An alternative way to pack tetrahedra is 
to introduce some edges in the structure that have six tetrahedra around them.
Such an arrangement is locally less favourable---for regular tetrahedra
there is an overlap equivalent to $63.17^\circ$. However, for
sufficiently large packings, the combination
of `gaps' and `overlaps' allows one to generate structures that have less
overall strain than polyicosahedral packings of the same size. 
For example, bulk polytetrahedral structures, called Frank-Kasper 
phases,\cite{FrankK58,FrankK59,Shoemaker} are now possible with these
two types of edge environments.
For clusters, such polytetrahedral
structures have a higher surface density than polyicosahedral
structures, and so need a smaller atomic size ratio to be stabilized in
binary systems.

Those edges that have six tetrahedra around them are said to
have disclination lines running along them, and 
such polytetrahedral structures can be 
can be viewed in terms of a network of disclination lines threading an 
icosahedrally-coordinated medium. Those atoms that have disclinations
passing through them have a coordination number greater than 12.
For example, 14-coordinate atoms have a single disclination line passing 
through them, and 15- and 16-coordinate atoms act as nodes for three and 
four disclination lines, respectively.
We will use the presence of these different types of Frank-Kasper 
coordination polyhedra to differentiate these polytetrahedral structures.

The ($\sigma_{BB}/\sigma_{AA}$,$N_A$) landscape associated with the 
BLJ$_{45}$ global minimum is shown in Fig.\ \ref{fig:BLJ45}.
It is noteworthy that the there is no optimal size ratio, but instead
the energy of the global minimum decreases virtually monotonically with
increasing size disparity. Also apparent is the magnitude of the 
stabilizations that can be achieved compared to monoatomic LJ clusters.
The lowest-energy structure in the $\sigma_{BB}/\sigma_{AA}$ range
that we consider here is $27.252\epsilon$ lower in 
energy than the LJ$_{45}$ global minimum, 
i.e.\ a 12.7\% decrease in the total energy.

There are four different compositional BLJ$_{45}$ global minima in 
the parameter range studied. Near to $\sigma_{BB}/\sigma_{AA}=1$, as one
would expect the global minimum is icosahedral with a Mackay overlayer, and
has the same geometric structure as for LJ clusters. The ideal composition
of this structure depends slightly on $\sigma_{BB}/\sigma_{AA}$. 
At $\sigma_{BB}/\sigma_{AA}=1.057$ the polytetrahedral structure A$_8$B$_{37}$
with a single negative disclination threading the structure becomes 
most stable.
At $\sigma_{BB}/\sigma_{AA}=1.071$ the core-shell anti-Mackay icosahedron
becomes most stable. 

These changes are driven by a balance between the number of 
nearest neighbours and the strain energy of the structures.
As $\sigma_{BB}/\sigma_{AA}$ increases, structures with a larger number 
of nearest neighbours become the most stable, when their strain energy is
sufficiently reduced.
In this series, the number of nearest neighbours increases from 180 to 
192 to 204, but at small $\sigma_{BB}/\sigma_{AA}$ so does the strain
energy also increase. In particular, notice that the strain energy 
for the polytetrahedral structure involving disclinations is less
than the polyicosahedral structure (Fig.\ \ref{fig:BLJ45}(b)).

These results illustrate how allowing two atoms of different sizes 
can substantially reduce the strain energy associated with 
polytetrahedral structures.
For example, the strain energy of the anti-Mackay icosahedron
can be decreased from 24.4 to $1.7\epsilon$ (Fig\ \ref{fig:BLJ45}e).
However, unlike BLJ$_{13}$, the strain cannot be completely removed
There are only two different nearest-neighbour distances in a 
13-atom icosahedron, whereas there are six 
different nearest-neighbour distances for the anti-Mackay icosahedron, 
and they cannot be adjusted so that they all have the 
same $r_{ij}/\sigma_{\alpha\beta}$.

When we also look at the compositional dependence of the global minimum in
Fig.\ \ref{fig:BLJ45}(b), we should remember that at 
$N_A=0$, $N_A=N$ and $\sigma_{BB}/\sigma_{AA}=1$
the model is the same as the one-component Lennard-Jones model.
Similarly, close to these values the structure has a Mackay overlayer, 
as for LJ clusters, i.e.\ for clusters with mostly A, mostly B or small 
$\sigma_{BB}/\sigma_{AA}$. By contrast, the polyicosahedral
structures are most stable for intermediate compositions
and larger $\sigma_{BB}/\sigma_{AA}$. For much of the structural phase 
diagram, sandwiched between these two structural types is a region where 
polytetrahedral structures with disclinations are most stable. However,
there is some asymmetry, as at larger $\sigma_{BB}/\sigma_{AA}$, this 
zone is only found for larger values of $N_A$. 

Our final case study is BLJ$_{55}$. This size has been chosen, as
it corresponds to the size at which a complete Mackay icosahedron 
is possible. Again, we illustrate the landscape associated with the
global minimum and, in this case, the seven compositional global minima 
(Fig.\ \ref{fig:BLJ55}). The overall behaviour is somewhat similar to 
BLJ$_{45}$. There are four main differences. Firstly,
the Mackay structures are most stable
over a greater range of $\sigma_{BB}/\sigma_{AA}$ and composition, which
is unsurprising given that $N=55$ is a Mackay magic number. 
Secondly, there is a more pronounced change in slope in Fig.\ 
\ref{fig:BLJ55}(a) associated with the onset of polytetrahedral structures.
The landscape is relatively flat in the regions where 
Mackay structures are most favoured, but
goes down more steeply, in the region where polytetrahedral structures
are most stable, because the strain energy associated with these structures
rapidly decreases, as the size disparity increases (Fig.\ \ref{fig:BLJ55}(e)).
Thirdly, polyicosahedral structures are most stable for a
smaller range of parameters. This is because, as the size increase the 
strain energy associated with these structures for the monoatomic LJ clusters 
increases rapidly, and hence the size difference that is needed to stabilize 
them also increases.
Fourthly, as a corollary of the above, the area in Fig.\ \ref{fig:BLJ55}(a) 
where polytetrahedral structures involving disclinations are more stable 
increases. Furthermore, there are two types of such structure, which
can be differentiated by the nodes in the disclination network.
The `$Z14$' structures involve a single disclination line passing
through the cluster, whereas the `$Z15$' structures has three disclinations
radiating out from the 15-coordinate atom.

For BLJ$_{55}$, the compositional global minimum changes
six times as $\sigma_{BB}/\sigma_{AA}$ increases. 
First, there is the core-shell Mackay icosahedron. 
Although the strain in this structure is much less than for polytetrahedral
structures, the strain energy can still be significantly reduced compared
to the LJ case by introduction of different-sized atoms 
(Fig.\ \ref{fig:BLJ55}(e)). For example,
this structure has its lowest energy at $\sigma_{BB}/\sigma_{AA}=1.064$. 
The next three compositional global minimum are polytetrahedral structures
that involve disclination lines. Note, that the $Z15$ structure is more stable
at smaller $\sigma_{BB}/\sigma_{AA}$ as it involves a greater disclination
density than the $Z14$ structures. These structures are not pure core-shell 
clusters, as although all the 12-coordinate interior atoms are A atoms, 
there is a preference for the larger atoms to lie on the disclinations, 
which is unsurprising given their higher coordination number.
The final three compositional global minima are polyicosahedral, and differ
just in the position of capping atoms and composition.

\subsection{\label{sect:vary_gmin}Compositional global minima}
\begin{figure}
\includegraphics[width=8.4cm]{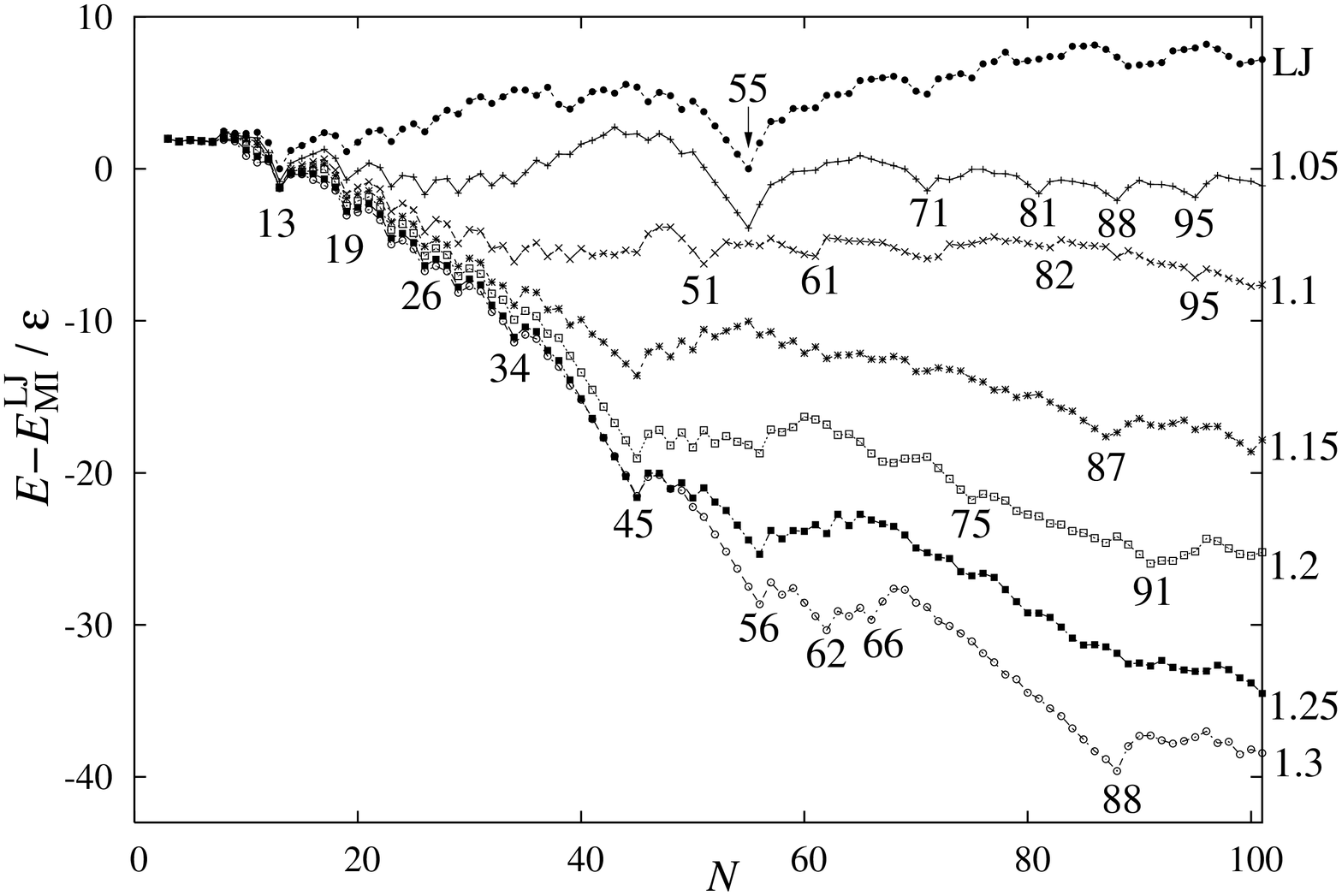}
\caption{\label{fig:eall}The energy of the BLJ global minimum for the
six values of $\sigma_{BB}/\sigma_{AA}$
studied, relative to $E^{\rm LJ}_{\rm MI}$, a fit 
to the energies of the Mackay icosahedra for LJ clusters.
A line corresponding to the LJ global minima is also included
}
\end{figure}

\begin{figure}
\includegraphics[width=8.4cm]{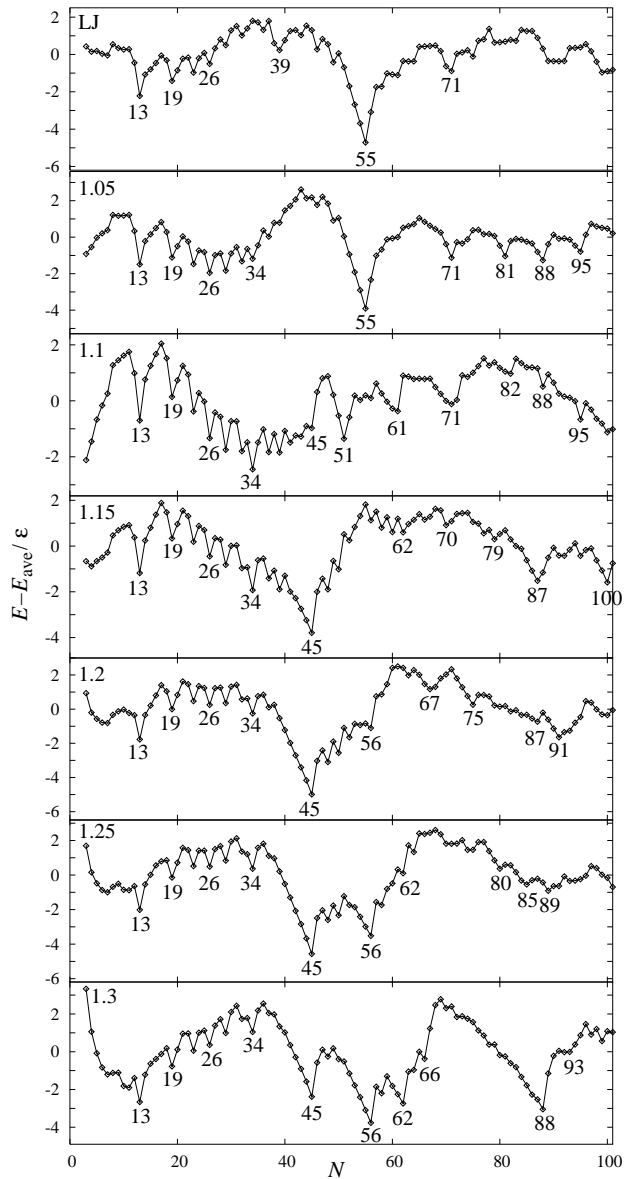}
\caption{\label{fig:enorm}The seven panels correspond to the
energies of the global minimum for the
six values of $\sigma_{BB}/\sigma_{AA}$
studied and for the one-component LJ clusters, 
relative to $E_{\rm ave}(\sigma_{BB}/\sigma_{AA})$, 
a fit to the energies of the global minima at that size ratio
using the form $a + b N^{1/3} + c N^{2/3} + d N$.}
\end{figure}

\begin{figure*}
\includegraphics[width=18cm]{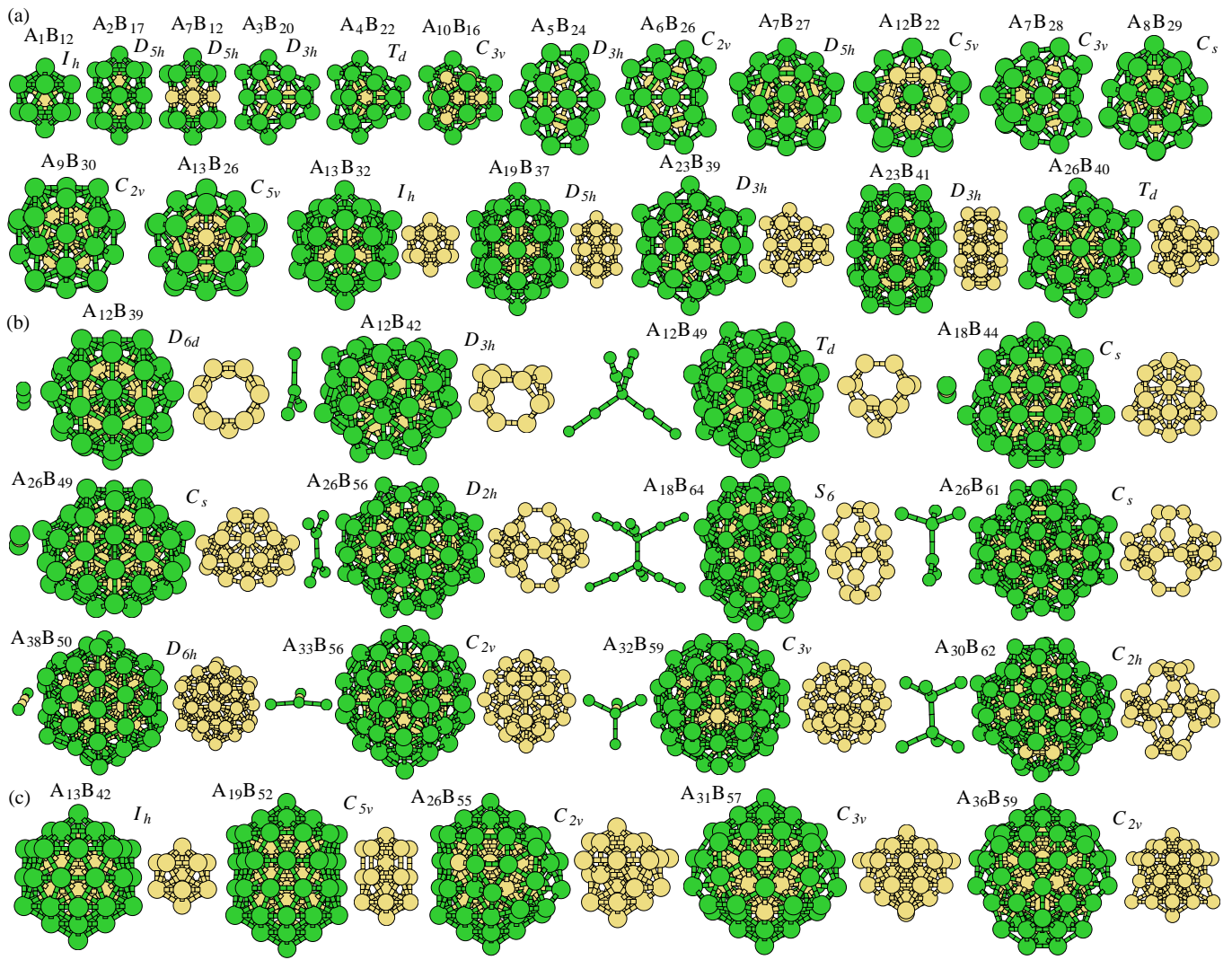}
\caption{\label{fig:struct} (Colour online)
A selection of the particularly stable BLJ global minima grouped
according to their structural type: (a) polyicosahedral, (b) polytetrahedral
with disclinations and (c) the 55-atom Mackay icosahedron with an anti-Mackay
overlayer.
For the larger clusters, to the right of the cluster, the A-atom core is also
depicted, and in (b) to the left is the disclination network.
}
\end{figure*}

\begin{figure}
\includegraphics[width=8.4cm]{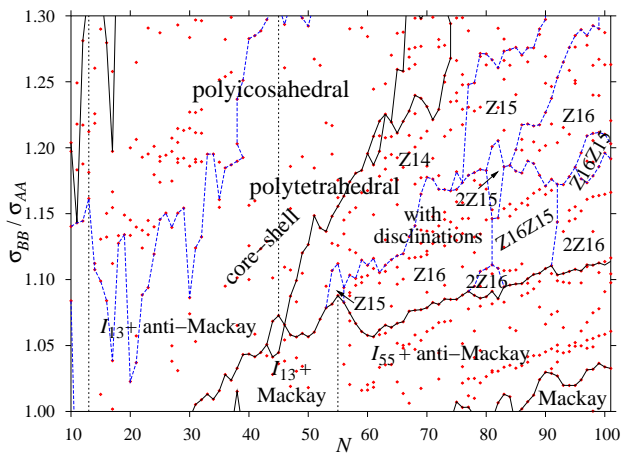}
\caption{\label{fig:phased} (Colour online)
Structural phase diagram showing how the structure
of the global minimum depends on $N$ and 
$\sigma_{BB}/\sigma_{AA}$. Each point corresponds to the
value of $\sigma_{BB}/\sigma_{AA}$ at which the global minimum
for a given $N$ changes. The lines divide the diagram into 
regions where the global minima have the same structural type. 
The labels $I_{13}$ and $I_{55}$ stand for the 13- and 55-atom 
Mackay icosahedra.}
\end{figure}

In this section, we focus our attention on the compositional global minima.
Figs.\ \ref{fig:eall}, \ref{fig:enorm}, \ref{fig:struct},  
and \ref{fig:phased}, summarize the results. 
The energies and points files for all the putative global minima are 
available online.\cite{CCDshort}
In Fig.\ \ref{fig:eall}, the energies at 
different size ratio are compared. This figure clearly illustrates that,
as noted in the above case studies, the
energy virtually monotonically decreases with increasing size ratio.
It also shows the substantial nature of the stabilizations that 
are achieved compared to the one-component Lennard-Jones system.

In Fig.\ \ref{fig:enorm}, the energies at the six size ratios we consider
are plotted in a way that reveals the particularly stable sizes. 
In Fig.\ \ref{fig:struct} we show some of the particularly interesting or 
stable structures associated with the different structural types.

Fig.\ \ref{fig:phased} provides a structural phase diagram showing how 
the ($N$,$\sigma_{BB}/\sigma_{AA}$) plane can be divided into regions
where the compositional global minima have the same type of structure.
To construct this diagram, we considered intervals of $0.01$ in 
$\sigma_{BB}/\sigma_{AA}$ and reoptimized the five lowest-energy structures 
from the nearest size ratios for which global optimization was performed. 
Then, we checked if the structure of the global minimum 
changes in any of these intervals, and if so we obtained the precise value of 
$\sigma_{BB}/\sigma_{AA}$ at which this change took place.
Of course, this approach will potentially miss the true
global minima at intermediate values of $\sigma_{BB}/\sigma_{AA}$ 
if it is not one of the five best for the values
at which we ran the global optimization algorithm, and it only allows 
for one change in the global minimum in any 0.01 interval. However, 
we are concerned more with the overall form of this diagram,
on which these approximations will only have a very minor effect.

The reference system to which to compare our results is, of course, 
the one-component LJ clusters, for which structures based on the
Mackay icosahedra are dominant in the current size range. 
Growth around the 13-atom icosahedron initially occurs in the 
anti-Mackay overlayer, but for LJ$_{31}$ and beyond
the global minimum has a Mackay overlayer.\cite{Northby87} Similarly, 
for the growth of the next icosahedral shell,
the global minima initially have an anti-Mackay overlayer, but for LJ$_{82}$ 
and beyond (with the exception of LJ$_{85}$) the Mackay overlayer is 
more stable.
The only exceptions to this dominance of icosahedral structures are for 
LJ$_{38}$, LJ$_{75-77}$ and LJ$_{98}$ for which an fcc truncated octahedron, 
Marks decahedra\cite{Doye95c} and a Leary tetrahedron,\cite{Leary99} 
respectively, are just more stable than the competing icosahedral structures.

As one moves away from $\sigma_{BB}/\sigma_{AA}=1$ in the structural phase 
diagram, firstly icosahedral structures quickly become
more stable than the five non-icosahedral structures mentioned above.
Secondly, for both the second and third icosahedral shells 
the crossover size at which the Mackay overlayer becomes more stable is pushed
to larger sizes (Fig.\ \ref{fig:phased}). 
This is because the anti-Mackay overlayer is more strained,
and so the introduction of two atomic sizes can lead to a greater reduction
in its strain energy. For both types of overlayer, the larger atoms will 
preferentially go into the surface layer, because of the tensile strain
in the surface of the icosahedral LJ clusters, leading to the formation
of core-shell clusters, as already illustrated by A$_{13}$B$_{32}$ 
and A$_{13}$B$_{42}$ in the previous section. 
There are many further examples in Fig. \ref{fig:struct}. 
As for the 13-atom icosahedron considered in the last section, there
is an optimal size ratio for these core-shell structures. Beyond this, 
compressive strains begin to build up in the surface until at some point
it becomes favourable to include some of the smaller A atoms in the 
surface. Some examples of the structures that result are shown in Fig.\
\ref{fig:struct}(a), 
e.g.\ A$_7$B$_{12}$, A$_{10}$B$_{16}$ and A$_{12}$B$_{22}$. 
However, the window of $\sigma_{BB}/\sigma_{AA}$ values for which these 
core-shell structures are most stable is wide, illustrating their particular 
stability.

This preference for the anti-Mackay overlayer can also be seen in the 
observed magic numbers (Fig.\ \ref{fig:enorm}). 
For the second shell, anti-Mackay clusters
are particularly stable at $N$=19, 23, 26 and 29 for LJ clusters, 
and correspond to 2, 3, 4 and 5 interpenetrating icosahedra. 
These become more prominent as $\sigma_{BB}/\sigma_{AA}$ increases with 
additional magic numbers at $N$=34 and 45, and to a lesser extent at 
$N$=32, 37 and 39. The magic number at $N=45$ is particularly 
strong as it corresponds to the completion of the anti-Mackay overlayer.
Interestingly, beyond $N=34$ the way the anti-Mackay 
overlayer grows depends on $\sigma_{BB}/\sigma_{AA}$. At smaller 
$\sigma_{BB}/\sigma_{AA}$ the formation of structures that 
can be thought of as interpenetrating complete icosahedra continues, 
as illustrated by A$_9$B$_{30}$.
In contrast, at larger $\sigma_{BB}/\sigma_{AA}$ the
sites above the faces of the central 13-atom icosahedra 
are completely filled and further growth occurs just by adding atoms 
above the vertices (Fig.\ \ref{fig:layer}), 
e.g.\ A$_{13}$B$_{26}$.

Similarly, for the third shell, as well as the 
magic number at $N$=71  already present for LJ clusters, additional
magic numbers appear at 81, 88 and 95. These correspond to covering 
five, eight, ten and twelve faces of the underlying Mackay icosahedron,
respectively.

As already mentioned, clusters with an anti-Mackay overlayer covering the
13-atom icosahedron are polyicosahedral in character. However, there is no
reason why this type of packing should not continue beyond the completion 
of this overlayer at $N=45$. Indeed, for the range of 
$\sigma_{BB}/\sigma_{AA}$ values considered here such structures are 
possible up until $N=69$, 
and the magic numbers at $N$=56, 62 and 66 that appear at larger 
$\sigma_{BB}/\sigma_{AA}$ correspond to core-shell 
polyicosahedral structures with the double, triple and quadruple 
icosahedra as their core (Fig.\ \ref{fig:struct}(a)). Also interesting
is the A$_{23}$B$_{41}$ structure, for which the core of A atoms is two 
face-sharing icosahedra.

In the structural phase diagram as $N$ increases
a region opens up between the Mackay
icosahedral and polyicosahedral structures,
where polytetrahedral structures involving disclinations are most stable.
The first such structure occurs at $N=44$ and as $N$ increases, these
structures occupying an increasing proportion of the phase diagram.
These structures are not seen at small size, firstly because the strain 
associated with polyicosahedral structures is not prohibitively high, and 
secondly because the introduction of disclinations would lead to 
high disclination densities, and unfavourable strain energies. For this
reason the Frank-Kasper coordination polyhedra that are possible at
$N$=15, 16 and 17 are never most stable.

Later when substantial strains have builit up in the polyicosahedral structures,
the introduction of disclinations along a small minority of the edges can lead
to a reduction in the overall strain energy. The structures 
A$_{12}$B$_{39}$, A$_{12}$B$_{42}$ and A$_{12}$B$_{49}$
provide interesting examples, and correspond to the covering of the
above Frank-Kasper coordination polyhedra by a complete anti-Mackay-like 
overlayer with atoms added above every face and vertex (except 
the six-fold vertices for the 51- and 54-atom structures).

At larger sizes more complex disclination networks are possible. For
example the $2Z16$ structures, e.g. A$_{18}$B$_{64}$, A$_{30}$B$_{62}$, 
have two nodes where four disclinations meet, and gives rise to an 
ethane-like disclination network, 
and the $Z15Z16$ structures, e.g. A$_{26}$B$_{61}$, 
have one node where three disclinations meet, and one where four meet.

From Fig.\ \ref{fig:phased} some systematic trends in the character of
the disclination networks are clear.
As $N$ increases, the structures next to the Mackay icosahedral-polytetrahedral
boundary have an increasing number of disclinations in order to reduce the 
growing strains that would have otherwise occurred.
However, as $\sigma_{BB}/\sigma_{AA}$ increases the different-sized atoms 
are able to reduce some of this strain, and the disclination density goes
down, until the boundary with disclination-free polyicosahedral structures 
is reached. So, for example, for BLJ$_{82}$ the structure changes from $2Z16$,
to $Z16Z15$ to $2Z15$ to $Z15$ to $Z14$ as $\sigma_{BB}/\sigma_{AA}$ increases.

\section{\label{sect:conc}Conclusions}

In this paper we have introduced a new global optimization approach for 
binary clusters that for a given size searches directly for the composition
of greatest stability. This development makes it tractable to explore
the size evolution of binary cluster structure systematically, rather than
just arbitrarily selecting particular sizes or compositions. 
Of course, by focussing on the compositional global minimum, some
information is lost, but as one is usually most interested in identifying 
the most stable magic number clusters, this is of little import. 
Besides the alternative approach of finding the global mininum at each size 
and composition would both be computationally extremely challenging and 
result in a surfeit of information, the added value of which is far from 
clear. 
The approach is straightforward to apply to other binary 
systems, and, for example, we have just completed applying this approach
to all clusters with up to 150 atom for a number of binary metal 
systems.\cite{JPKDRFinprep}

One of the most important aims of theoretical studies of structure is 
to provide models that can aid the interpretation of experimental observations.
In this regard, it is reasonable to ask whether our approach
of optimizing the composition 
is realistic of what might be occurring experimentally.
The answer would of course depend on how the binary clusters
are produced experimentally, but recent experiments on copper-tin
clusters showed that 
the particularly stable compositions could be obtained after annealing the 
clusters.\cite{Breaux05d}

We have used the current approach to systematically explore the structural 
effects of having different-sized atoms on binary LJ clusters. 
Particularly interesting is how this stabilizes polytetrahedral clusters,
both those that are polyicosahedral and those that involve disclinations,
which for the one-component LJ clusters are too strained to be competitive.
The effects of the two atom sizes is somewhat similar to other methods of 
strain relief, such as widening the potential well, which also pushes the 
anti-Mackay to Mackay transitions to larger size and stabilizes 
polytetrahedral structures with disclination lines.\cite{Doye95c,Doye97d} 
In future work, we plan to extend our exploration of the BLJ system
to other choices for the four parameters in the potential, and to help
develop its status as an archetypal model system for which to 
understand the structure of binary clusters.

Our results have provided a zoo of interesting structures, which 
are good candidates for particularly stable clusters for binary systems
where the two atom types have significant differences in size.
Although BLJ clusters are not a realistic model for much, except for rare 
gas mixtures at certain parameter choices, as for monoatomic LJ clusters,
our expectation is that the stable structural forms seen for the BLJ clusters
are robust, and likely to be relevant for a wide variety of systems, where
the interactions are approximately isotropic. 
This confidence is based on previous experience, where particular 
structural forms first seen for a model system are later found in real systems.
For example, the Leary tetrahedron,\cite{Leary99} which was first discovered 
as the global minimum for LJ$_{98}$, was later found to be one of 
the dominant structural forms for clusters of C$_{60}$ molecules.\cite{Branz00}
Indeed, core-shell polyicosahedral structures have already been found for 
binary metal clusters, such as the Ag-Ni and Ag-Cu systems.\cite{Rossi04,Rapallo05,Ferrando05,JPKDRFinprep}  

In this paper, we have focussed on the global minima of our system, 
and so Fig.\ \ref{fig:phased} represents the structural phase diagram
at zero temperature. It is, therefore legitimate to ask how our results
would be affected by temperature. Firstly, for sufficiently small 
size differences between the atoms, it is likely that there will be 
low-temperature order-disorder transitions, where the geometric structure is 
retained, but the permutational order is lost.\cite{Lopez93,Frantz97} 
However, as $\sigma_{BB}/\sigma_{AA}$ increases, these transitions 
are likely to become less common, because the A and B atoms quickly 
develop strongly preferred positions, as, for example, in the core-shell 
ordering.

Secondly, it is known for the LJ system that an
effect of temperature is to push the anti-Mackay to Mackay transition to 
larger size,\cite{Frantz01,Calvo01a,Frantsuzov05,Noya06b} because of the 
greater vibrational entropy of the anti-Mackay structures.\cite{Doye02b}
Indeed, the form of Fig.\ \ref{fig:phased} at small $\sigma_{BB}/\sigma_{AA}$ 
looks similar to the $(N,T)$ structural phase diagram for LJ 
clusters.\cite{Frantz01,Frantsuzov05} 
Therefore, temperature will reinforce the stabilization of polytetrahedral
structures seen for BLJ clusters.

Here, we have probed all BLJ clusters with up to 100 atoms. But what would
one expect at larger sizes, and in the bulk limit? The dominance of 
polytetrahedral structures in the current size range is because of their
favourable surface energetics, and it can be seen in Fig.\ \ref{fig:phased}
that the range of stability for non-polytetrahedral clusters is increasing 
as $N$ increases. 
The latter would suggest that Frank-Kasper phases are never the ground state
in the bulk limit. Instead, phase-separated A and B fcc crystals are likely
to have the lowest energy for bulk. However, one cannot rule out that this 
boundary might flatten off at larger sizes, nor that there might be 
some range of temperature and pressure, where Frank-Kasper phases might 
be most stable. 

The reverse of the above prediction is that there are likely to be binary
systems, which although they do not have any stable bulk 
Frank-Kasper phases, nevertheless exhibit Frank-Kasper-like structures
for their clusters. Indeed, we have already identified such structures
for Ni-Al, Ag-Cu and Ag-Ni clusters using the current 
methodology.\cite{JPKDRFinprep}
Although likely to be less common than for binary clusters, there are a few 
one-component systems that also exhibit these
polytetrahedral clusters.\cite{Doye97d,Doye01d,Doye03d}

Our results are also of relevance for understanding the effect of 
size mismatch on glass formation.
In order to explain the difficulty of homogeneous nucleation in supercooled
metal droplets, Frank argued that if the preferred local order within the
liquid was incompatible with the local structure in the crystal, 
nucleation would be suppressed because of the substantial structural 
rearrangement required.\cite{Frank52} 
Furthermore, he used the structures of isolated
LJ clusters, in particular the stability of the 13-atom icosahedron compared
to an fcc cluster of the same size, to provide a picture of the local
order within the liquid. More recently, the predictive power of isolated 
clusters for understanding liquid structure has received 
significant empirical support.\cite{Doye01a,Dzugutov02b,Doye03b}

Empirically, it has been found that one of the conditions
for metallic alloys to form bulk glasses is that there is a size difference
of at least 12\%.\cite{Inoue00} A simple theoretical justification of this
can be given in terms of the size difference required to destabilize
a crystalline solid solution.\cite{Egami84,Egami97} For the current system,
the critical value of $\sigma_{BB}/\sigma_{AA}$ at which a solid solution
is no longer always stable has been found to be 1.20.\cite{Li93}
Although the determinants of glass-forming ability are 
subtle,\cite{Fernandez04,Jalali05}
and more sophisticated theories can be 
developed,\cite{Senkov01,Miracle03,Senkov05} our results highlight the 
potential role played by the local structure within the 
liquid,\cite{Fernandez04b} and how size differences can stabilize local 
polytetrahedral, in particular icosahedral,\cite{Lee03} order,
and hence frustrate crystallization.
Indeed, one of the commonly used model glass-forming systems used in
the simulation community is a 50:50 BLJ mixture with 
$\sigma_{BB}/\sigma_{AA}=1.2$, and all other parameters as in the present
study.\cite{Wahnstrom91,Schroder00,Lacevic03}

\begin{acknowledgments}
J.P.K.D is grateful to the Royal Society for financial support.
\end{acknowledgments}


\begin{thebibliography}{75}
\bibitem[{pre()}]{prevdef}
\bibinfo{note}{Previously, we have defined the strain energy with respect to
  the equilibrium pair separation,\cite{Doye95c} rather than the average
  nearest-neighbour distance. The problem with this previous definition is that
  essentially unstrained structures normally have an average neighbour distance
  that is smaller than the equilibrium pair separation, because this decreases
  $E_{\rm nnn}$ at the expense of a smaller increase in $E_{\rm nn}$.}
\end{thebibliography}
\end{document}